\documentclass[10pt]{article}
\usepackage{graphicx,longtable,mathrsfs,color,array}
\usepackage[hidelinks]{hyperref}
\usepackage{amsmath}
\usepackage[left=2cm, right=2cm, top=2cm, bottom=2cm]{geometry}
\graphicspath{{figures/}}
\usepackage{titlesec}
\usepackage{mathrsfs}
\titleformat*{\section}{\large\bfseries}
\titleformat*{\subsection}{\normalsize\bfseries}

\begin{document}

\begin{center}
{\large \bf Continuity equations for general matter: applications in numerical relativity} \\
\vspace{0.5cm}
\small Katy Clough
\footnote{katy.clough@physics.ox.ac.uk} \\
Astrophysics, University of Oxford, DWB, Keble Road, Oxford OX1 3RH, UK
\end{center}

\begin{abstract}
Due to the absence of symmetries under time and spatial translations in a general curved spacetime, the energy and momentum of matter is not conserved as it is in flat space. This means, for example, that the flux of matter energy through a surface is in general not balanced by an equal increase in the energy of the matter contained within the enclosed volume - there is an additional ``source'' resulting from the curvature of spacetime acting on the matter (and vice versa). One can calculate this source term and reconcile the flux and energy accumulation over time in an arbitrary volume, although a foliation of the spacetime must be chosen, making the quantities inherently coordinate dependent. Despite this dependence, these quantities are practically useful in numerical relativity simulations for a number of reasons. We provide expressions for general matter sources in a form appropriate for implementation in the ADM decomposition, and discuss several applications in simulations of compact object dynamics and cosmology.
\end{abstract}

\section{Introduction}

In a curved spacetime, the flux of matter energy-momentum through a surface is in general not balanced by an equal increase in the energy and momentum of the matter contained within the enclosed volume. In the Newtonian picture this mismatch is thought of as resulting from the gravitational force acting on the fluid, but in a general relativistic description no such force exists, and instead the change is a result of the transfer of energy and momentum between the matter and the spacetime curvature. This latter description captures effects that cannot easily be described by Newtonian forces - a good example is a de Sitter expansion in FLRW cosmology, in which the total matter energy in a comoving volume grows over time with the expansion, despite no flux crossing the surface; in a flat space view, energy is created ``out of nothing''.

Describing this transfer of energy between curvature and matter is inherently problematic because of the difficulty of defining an energy for the gravitational field at some specific event; since the field is locally undetectable, its energy must be zero at any point in local inertial coordinates. One can get around this by considering integrated quantities over a finite spatial volume, but in doing so one has to choose a slicing of the 4 dimensional spacetime, and thus the description becomes inherently gauge dependent. Setting aside the special case of spherical symmetry, the only thing that can be unambiguously defined is the total ADM mass and momenta of the spacetime at spatial infinity (assuming asymptotically flatness); the division of this total energy between the spacetime and the matter unavoidably depends on the coordinates chosen to describe the non asymptotically flat region.

Whilst acknowledging these difficulties in interpretation, it is nevertheless useful to define and inspect such quantities during numerical relativity (NR) simulations, or strong gravity simulations more generally, such as those with fixed background metrics. In particular, given the numerical slicing of the spacetime into spatial hypersurfaces, it is possible to reconcile the instantaneous fluxes and energy accumulation of the matter in an arbitrary spatial volume by an application of Gauss's law with an additional ``source'' term that quantifies the transfer of energy or momentum from the spacetime curvature. The expressions so obtained are essentially the continuity equation and Euler equations for the conservation of the energy-momentum of the matter, with the effect of curvature treated as a source term.

The calculation of this additional source of energy-momentum from curvature is useful in numerical simulations of strong gravity environments for a number of practical reasons. Primarily, it provides a very useful check on the consistency and correctness of the calculation of the flux and energy accumulation - where the spacetime is not well resolved the quantities will not match up, and their disagreement is a useful measure of the non-convergence of the simulation as a whole. This is particularly important in simulations performed in the decoupling limit where matter is evolved on a fixed or dynamical metric for which the matter backreaction is neglected, because here one cannot check the satisfaction of the Hamiltonian and Momentum constraints. Secondly, the source term is a volume integral and in situations where the system reaches a steady state, it tends to converge faster to its final value, and be less prone to boundary issues than the surface fluxes, making it easier to measure in time limited simulations (we will give an example of this in the context of dynamical friction simulations). Finally, it provides an intuitive quantification (albeit a gauge dependent one) of how the curvature and matter are interacting at different events in the spacetime, which can aid one's physical understanding of the evolution in complex dynamical cases. 

The purpose of this note is to provide the relevant expressions for these sources for general matter in a format suitable for implementation in NR codes, and discuss applications and potential issues. 

The expressions presented here are essentially the same as the energy-momentum conservation expressions that form the basis of the formulation of conservative hydrodynamical variables in NR codes (see any of the standard NR texts, e.g. section 7.3 of Alcubierre \cite{alcubierre2008introduction}, section 5.2 of Baumgarte and Shapiro \cite{baumgarte_shapiro_2010} and section 4.3 of Shibata \cite{ShibataBook}), where satisfying the relevant balance laws is built into the formulation. However, the expressions here are for a general source, and so, for example, they can be applied to real scalar fields and other types of matter with no natural flux-conservative form. In such cases it is relatively uncommon to see the relevant balance law being checked, except in cases with fixed background spacetimes containing Killing vectors (see e.g. \cite{East:2017mrj, Bamber:2020bpu}), where the sources are identically zero.

The fundamental ideas around conservation in a curved spacetime date back to the Einstein's gravitational pseudotensor \cite{Einstein}, and Landau and Lifshitz's alternative pseudotensor \cite{Landau:101808}. These ideas are thoroughly explained in Chapters 19 and 20 of Gravitation by Misner Thorne and Wheeler (MTW) \cite{misner1973gravitation}, and \cite{DeHaro:2021gdv} provides an interesting recent review of their philosophical difficulties.

The note is structured as follows:
\begin{itemize}
	\item In Sec. \ref{sec-curved} we start by reviewing the derivation of conserved quantities in a spacetime with a Killing vector, using the 3+1 ADM picture.
	\item In Sec. \ref{sec-noncon} we cover the case where the vector direction defining the current is no longer a symmetry of the spacetime and give expressions for the additional source terms obtained for time-like (energy) and spatial (linear momentum) directions in terms of the ADM quantities.
	\item In Sec. \ref{sec-gauge} we discuss gauge dependence and relate the source to observed quantities. We emphasise in particular that the continuity equations derived focusses on the stress-energy of the \textit{matter} and not of the \textit{spacetime} so additional ingredients are required to infer, for example, the ADM mass or momenta of the system as a whole.
	\item In the final section, Sec. \ref{sec-applications} we will discuss three simple illustrative examples that give some additional physical insight to the role of the source: dynamical friction on a black hole (BH), real scalar field solitons (``oscillatons''), and FLRW cosmology.
\end{itemize}
A more detailed discussion of the application of Gauss's Law to the 4D ADM spacetime is given in the Appendix.

\section{Continuity in a curved spacetime with Killing vectors}
\label{sec-curved}

Our notations follow the typical conventions in NR, in particular, the units are geometric, $G=c=1$ and we use the ADM (Arnowitt Deser Misner) decomposition \cite{Arnowitt:1959ah}\footnote{For those new to NR a very user-friendly reference is \cite{baumgarte_shapiro_2021}, or Gourgoulhon \cite{Gourgoulhon:2007ue} gives a more thorough but still very readable treatment.}, in which the metric $g_{\mu\nu}$ is expressed in adapted coordinates in terms of a spatial metric $\gamma_{ij}$, a lapse $\alpha$ and a shift vector $\beta^i$ as
\begin{equation}
	ds^2 = - (\alpha^2 - \beta^i \beta_i) dt^2 + 2 \beta_i dx^i dt + \gamma_{ij} dx^i dx^j ~, 
\end{equation}
and the extrinsic curvature $K_{ij}$ is related to the Lie derivative along the normal direction to the spatial slice $n_\mu = (- \alpha, 0, 0, 0)$ as
\begin{equation}
	K_{ij} = - \frac{1}{2} \mathcal{L}_{\hat{n}} \gamma_{ij} = \frac{1}{2\alpha} (-\partial_t \gamma_{ij} + D_i \beta_j + D_j \beta_i) ~.
\end{equation}

We express the stress energy tensor $T_{\mu\nu}$ in terms of the quantities measured by the normal observers, energy density $\rho$, momentum density $S_i$ and stress energy density $S_{ij}$, which are those typically calculated in NR codes
\begin{equation}
	\rho = n^\mu n^\nu T_{\mu\nu} \quad S_i = - n^\mu T_{\mu i} \quad S_{ij} = T_{ij} ~.
\end{equation}
From these the full stress energy tensor can be reconstructed as
\begin{equation}
	T^{\mu\nu} = \rho n^\mu n^\nu + S^\mu n^\nu + S^\nu n^\mu + S^{\mu\nu} ~,
\end{equation}
noting that in the adapted coordinate basis, $S^t = S^{\mu t} = 0$ (whilst in general $S_t \neq 0$ and $S_{\mu t} \neq 0$).

We define a current $J^\mu$ by projecting the stress energy tensor $T_{\mu\nu}$ in the direction $\zeta^\mu$
\begin{equation}
	J^\mu = \zeta^\nu T^\mu_\nu ~.
\end{equation}
If $\zeta^\nu$ is a Killing vector, $\mathcal{L}_{\vec{\zeta}} g_{\mu\nu} = 0$, then $\nabla_\mu J^\mu=0$ and the current is conserved throughout the spacetime in the usual ``flat space'' sense that the energy accumulated in some volume is balanced by the flux through its surface. Explicitly we have 
\begin{equation}
	\int d^4 x ~ \sqrt{-g} ~ \nabla_\mu J^\mu = 0 ~,
\end{equation}
from which we can convert the 4D volume integral to a 3D surface integral according to Gauss Law. The setup as applied to the ADM slicing is discussed in more detail in the Appendix, but in summary one finds that on any 3D spatial hypersurface $\Sigma$ of constant time coordinate $t$
\begin{equation}
	\partial_t \int_{\Sigma} d^3 x \sqrt{\gamma} ~ Q = - \int_{\partial\Sigma} d^2 x \sqrt{\sigma} ~ F ~.
\end{equation}
Here we have identified a charge $Q$
\begin{equation}
	Q = - n_\mu J^\mu = - n_\mu \zeta^\nu T^\mu_\nu 
\end{equation}
and a flux $F$ out of the spatial boundary $\partial\Sigma$
\begin{equation}
    F = \alpha N_i J^i = \alpha N_i T^i_\nu \zeta^\nu ~.
\end{equation}
For an isolated system with zero net flux through the surface, the total amount of the charge $Q$ is conserved.
Here $\gamma$ is the determinant of the spatial metric, $N^i$ is the outward normal direction to the 2D bounding surface $\partial \Sigma$ on which the induced metric has determinant $\sigma$. $N^i$ is normalised such that $\gamma_{ij}N^i N^j = 1$, and $n^\mu$ is the normal to $\Sigma$ as defined above for which $g_{\mu\nu} n^\mu n^\nu = -1$.
\footnote{Note that whilst writing the integrals in this way makes more sense physically, in a practical implementation it is not usually necessary to calculate $\sigma$ and $N^i$ explicitly, since one can directly use the coordinate vector $s^i$ and the fact that $\sqrt{\sigma} N_i = \sqrt{\gamma} s_i$ - see the Appendix for more details.}

Let us first consider energy conservation, and choose to define the energy-current as the flux in the time-like direction defined by $\zeta^\mu = t^\mu = (1,0,0,0)$, which is a Killing vector when $\partial_t {g_{\mu\nu}}=0$, for example in the Kerr metric.
Note that we could instead use the time-like direction that defines the surfaces on which we calculate the total conserved charge at some time $t$, that is $n^\mu$ - that of the normal observers. In numerical simulations, $n^\mu$ and $t^\mu$ are usually distinct for singularity avoidance (a non-constant lapse) and to mitigate the resultant slice-stretching (a non-zero shift). In a gauge with a non zero shift (e.g. the puncture gauge that is commonly used for black holes spacetimes) the normal direction does not correspond to a Killing direction once a stationary state is reached, whereas $t^{\mu}$ does. However, in a dynamical spacetime this is an arbitrary choice, and we will see later that the use of either $\zeta^\mu = t^\mu = (1,0,0,0)$ or $\zeta_\mu = n_\mu/ \alpha = (1,0,0,0)$ to define the current is related to using either the Einstein or Landau Lifshitz pseudotensors when defining the conserved gravitational stress energy tensor.
For now, taking the case $\zeta^\mu = t^\mu = (1,0,0,0)$, we have
\begin{equation}
	Q = Q_t = - n_\mu J^\mu = - n_\mu \zeta^\nu T^\mu_\nu = \alpha T^0_0 = - \alpha \rho + \beta_k S^k \label{eq:rhoT}
\end{equation}
and
\begin{equation}
    F = F_t = \alpha N_i J^i = \alpha N_i T^i_0 
    = N_i \left( \beta^i (\alpha \rho - \beta^j S_j) + \alpha (\beta^k S^i_k - \alpha S^i)  \right) ~.
\end{equation}
Using these expressions, we should be able to equate the time derivative of the volume integral of $Q_t$ with (minus) the surface integral of the flux $F_t$. 

One can also consider the case of a spatial Killing vector $\zeta^\mu = \delta^\mu_i$, representing the conservation of momentum (where for example $i=x$ for linear momentum in the $x$ direction). Now we have momentum charge
\begin{equation}
	Q = Q_i = \alpha T^0_i = S_i ~,
\end{equation}
and the flux is the $i$-stress in the direction normal to the surface $\partial \Sigma$ (aka the $i$-momentum flux through the surface)
\begin{equation}
    F = F_i = \alpha N_j T^j_i = N_j ( \alpha S_i^j - \beta^j S_i ) ~.
\end{equation}
Such expressions are commonly used as diagnostics in simulations of matter on a fixed Kerr or Schwarzschild black hole background, e.g. in superradiance calculations \cite{East:2017mrj}, or accretion onto black holes \cite{Bamber:2020bpu}. Note that such calculations are always approximations - in a spacetime with a truly time independent metric, the matter must have also reached a stationary state, and so evolving matter on a fixed background is a fundamentally inconsistent thing to do (but often a very good approximation for small matter densities, as discussed in Sec. \ref{sec-applications}).

\section{Continuity in a general curved spacetime}
\label{sec-noncon}

Now let's consider the case where the current is defined using a direction $\tilde \zeta^\mu$ that is \textit{not} a Killing vector. 

Using the same procedure as in the previous section, we start with the 4D integral of the divergence, which is no longer equal to zero but instead
\begin{equation}
	\int d^4 x ~ \sqrt{-g} ~ \nabla_\mu J^\mu = \int d^4 x ~ \sqrt{-g} ~ T^\mu_\nu \nabla_\mu \tilde{\zeta}^\nu ~,
\end{equation}
from which we obtain
\begin{equation}
	\partial_t \left( \int_{\Sigma} d^3 x \sqrt{\gamma} ~ Q \right) = - \int_{\partial\Sigma} d^2 x \sqrt{\sigma} ~ F + \int_\Sigma d^3 x \sqrt{\gamma} ~ \mathcal{S} ~.
	\label{eq:nosingularity}
\end{equation}
Here in addition to the charge and the flux, which remain the same as in the previous section, we now have an additional ``source'' $\mathcal{S}$
\begin{equation}
    \mathcal{S} = \alpha T^\mu_\nu ~\nabla_\mu \tilde{\zeta}^\nu \label{eq:source} ~.
\end{equation}
Although the source is a scalar quantity, it relates to a particular set of observers for whom $Q$ is the measured charge (in the case in which $Q$ is the energy, for example, $\tilde{\zeta}^\mu$ is the direction they call time).
In most numerical situations one will be interested in quantities measured by coordinate observers, such that $\tilde{\zeta}^\mu$ is one of the coordinate basis vectors (with components like (1,0,0,0) for time), then $\partial_\mu \tilde{\zeta}^\nu = 0$ and we can write
\begin{equation}
    \mathcal{S} =  \alpha T^\mu_\nu ~^{(4)}\Gamma^\nu_{\mu \rho} \tilde{\zeta}^\rho \label{eq:source2} ~.
\end{equation}
This expression is calculated in a specific frame, and is manifestly coordinate dependent as one can see from the presence of the Christoffel symbol.  Whilst Eq. \eqref{eq:source2} covers the cases below, Eq. \eqref{eq:source} is more general and would be required for more complicated scenarios (such as the treatment of angular momentum in Cartesian coordinates).

Physically, we can see from Eq. \eqref{eq:source} that the source describes how the curvature of the spacetime guides the flow of energy-momentum in the fluid from other directions into (or away from) energy-momentum in the $\tilde{\zeta}^\mu$ direction, thus breaking the conservation of the charge that we would have in flat space (or one possessing the appropriate Killing vector). This interpretation will be made more explicit in Sec. \ref{sec-gauge}.

In the case where we choose $\tilde{\zeta}^\mu = \delta^\mu_t$ we are considering the flux and accumulation of energy in the spacetime, as measured by the coordinate observers. Here the source is
\footnote{The expressions for the 4D Christoffel symbols in terms of the ADM quantities can be found in Appendix B of \cite{alcubierre2008introduction}.}
\begin{equation}
    \mathcal{S}_t = \alpha T^\mu_\nu ~^{(4)}\Gamma^\nu_{\mu t} = - \rho \partial_t \alpha 
           + S_i \partial_t \beta^i  +  \frac{\alpha}{2} S^{ij} \partial_t \gamma_{ij} ~.
           \label{eq:Esource}
\end{equation}

In the case of a spatial $\tilde{\zeta}^\mu = \delta^\mu_i$ we are considering the flux and accumulation of $i$-momentum in the spacetime. Here we obtain
\begin{equation}
    \mathcal{S}_i = \alpha T^\mu_\nu ~^{(4)}\Gamma^\nu_{\mu i} = - \rho \partial_i \alpha + S_j \partial_i \beta^j + \alpha S^k_j ~^{(3)}\Gamma^j_{ik} ~. \label{eq:Psource}
\end{equation}
In both cases one can see that the sources are zero in a spacetime with the appropriate Killing vector.

We should therefore be able to reconcile the flux across some surface with the rate of change of the charge in the enclosed volume, plus the volume integral of the source term, for any type of matter.
This is an instantaneous expression obeyed for any volume at any time $t$ and so it will be true in any spacetime, not just ones with a static background or those where the outer surface is taken to be in a region that is asymptotically flat (although in such general cases the components may be harder to interpret physically, as discussed further in Sec. \ref{sec-gauge}). Therefore one can integrate the quantities over time during a simulation to confirm that the above expression remains satisfied. This is usually a very good check of whether the different scales of interest are appropriately resolved.

When there is a singularity in the spacetime, we can no longer perform a volume integral over the whole spatial slice, and we therefore need to replace a volume containing the singularity $\Sigma_i$ with the flux at some inner surface $\partial\Sigma_i$ enclosing it, i.e.
\begin{equation}
	\partial_t \left( \int_{\Sigma - \Sigma_{i}} d^3 x \sqrt{\gamma} ~ Q \right) = - \left( \int_{\partial\Sigma} d^2 x \sqrt{\sigma} ~ F - \int_{\partial\Sigma_{i}} d^2 x \sqrt{\sigma} ~ F \right) + \int_{\Sigma - \Sigma_{i}} d^3 x \sqrt{\gamma} ~ \mathcal{S} ~.  
	\label{eq:general}
\end{equation}
$\Sigma_i$ is often chosen to be at or near the horizon, which allows us to relate the measured flux to energy or momentum accretion onto the BH, but this quantity is highly coordinate dependent, and in practise a slightly larger radius can be used (and may be required for a Schwarzschild type gauge where the horizon is a coordinate singularity).

\section{Gauge dependence and relation to total energy-momentum of the spacetime}
\label{sec-gauge}

Having worked out the effect of curvature on the matter content, we now describe how this is related to the total energy and momentum of the spacetime. 

Einstein's pseudotensor for the gravitational field energy-momentum $t^\mu_\nu$ is defined as
\begin{equation}
{t_{{\mu }}}^{{\nu }}={\frac  {1}{16\pi{\sqrt  {-g}}}}(\partial_{\mu }(g^{{\alpha \beta }}{\sqrt  {-g}})(\Gamma _{{\alpha \beta }}^{{\nu }}-\delta _{{\beta }}^{{\nu }}\Gamma _{{\alpha \sigma }}^{{\sigma }})-\delta _{{\mu }}^{{\nu }}g^{{\alpha \beta }}(\Gamma _{{\alpha \beta }}^{{\sigma }}\Gamma _{{\sigma \rho }}^{{\rho }}-\Gamma _{{\alpha \sigma }}^{{\rho }}\Gamma _{{\beta \rho }}^{{\sigma }}){\sqrt  {-g}}) ~, \label{eq-einsteinpseudo}
\end{equation}
which can be shown to satisfy the conservation law
\begin{equation}
	\partial_\mu (\sqrt{-g} (T^\mu_\nu + t^\mu_\nu) ) = 0 ~.
\end{equation}
The pseudotensor $t^\mu_\nu$ is a second order quantity in the first derivatives of $g_{\mu\nu}$, and so whilst in local inertial coordinates it will reduce to zero, its integral over a volume is directly related to the energy momentum of the spacetime curvature.
One can show that in an asymptotically flat space the ADM mass and momenta $P_\mu$ agree at infinity to the volume integral of the relevant charge, ie,
\begin{equation}
	P_\mu = \lim_{r\to\infty}\int d^3 x \sqrt{-g}(T_\mu^0 + t_\mu^0) ~,
\end{equation}
assuming the usual conditions regarding asymptotically Minkowski coordinates for the ADM quantities are met (see Gourgoulhon section 8.2.1 \cite{Gourgoulhon:2007ue}).

We have calculated the matter half of this conservation law, that is, (assuming that $\tilde{\zeta}^\mu$ is the $\nu$ coordinate direction) our source is simply
\begin{equation}
	\sqrt{\gamma} ~ \mathcal{S}_\nu = \sqrt{-g} ~T^\mu_\rho ~^{(4)}\Gamma^\rho_{\mu \nu} = \partial_\mu (\sqrt{-g} T^\mu_\rho \delta^\rho_\nu) = \partial_\mu (\sqrt{-g} T^\mu_\nu)  ~.
\end{equation}
and therefore 
\begin{equation}
	\sqrt{\gamma} ~ \mathcal{S}_\nu =  -  \partial_\mu (\sqrt{-g} t^\mu_\nu)  ~.
\end{equation}
Thus what we have called the source $\mathcal{S}_\nu$ does indeed tell us the transfer of energy and momentum between the source and the curvature within the volume, in the framework of the Einstein pseudotensor. 

Note that to infer from this the \textit{total} energy or momentum of the spacetime we need further information - specifically, we need to know the initial volume integral of the curvature charge and any curvature fluxes (i.e., gravitational waves) that exit the volume under consideration. If this is required, one could add the relevant expressions from Eq. \eqref{eq-einsteinpseudo} into the simulation, or, if the outer boundary of the volume lies in an asymptotically flat region, recover them from the gravitational wave flux and deficit in the total ADM mass at this surface.

An alternative is provided by the Landau Lifshitz pseudotensor $t_{LL}^{\mu\nu}$
\begin{multline}
t_{{LL}}^{{\mu \nu }}={\frac  {1}{16\pi}}((2\Gamma _{{\alpha \beta }}^{{\sigma }}\Gamma _{{\sigma \rho }}^{{\rho }}-\Gamma _{{\alpha \rho }}^{{\sigma }}\Gamma _{{\beta \sigma }}^{{\rho }}-\Gamma _{{\alpha \sigma }}^{{\sigma }}\Gamma _{{\beta \rho }}^{{\rho }})(g^{{\mu \alpha }}g^{{\nu \beta }}-g^{{\mu \nu }}g^{{\alpha \beta }}) \\
{\displaystyle +g^{\mu \alpha }g^{\beta \sigma }(\Gamma _{\alpha \rho }^{\nu }\Gamma _{\beta \sigma }^{\rho }+\Gamma _{\beta \sigma }^{\nu }\Gamma _{\alpha \rho }^{\rho }-\Gamma _{\sigma \rho }^{\nu }\Gamma _{\alpha \beta }^{\rho }-\Gamma _{\alpha \beta }^{\nu }\Gamma _{\sigma \rho }^{\rho })}+g^{{\mu \alpha }}g^{{\beta \sigma }}(\Gamma _{{\alpha \rho }}^{{\nu }}\Gamma _{{\beta \sigma }}^{{\rho }}+\Gamma _{{\beta \sigma }}^{{\nu }}\Gamma _{{\alpha \rho }}^{{\rho }}-\Gamma _{{\sigma \rho }}^{{\nu }}\Gamma _{{\alpha \beta }}^{{\rho }}-\Gamma _{{\alpha \beta }}^{{\nu }}\Gamma _{{\sigma \rho }}^{{\rho }}) \\
{\displaystyle +g^{\nu \alpha }g^{\beta \sigma }(\Gamma _{\alpha \rho }^{\mu }\Gamma _{\beta \sigma }^{\rho }+\Gamma _{\beta \sigma }^{\mu }\Gamma _{\alpha \rho }^{\rho }-\Gamma _{\sigma \rho }^{\mu }\Gamma _{\alpha \beta }^{\rho }-\Gamma _{\alpha \beta }^{\mu }\Gamma _{\sigma \rho }^{\rho })}+g^{{\nu \alpha }}g^{{\beta \sigma }}(\Gamma _{{\alpha \rho }}^{{\mu }}\Gamma _{{\beta \sigma }}^{{\rho }}+\Gamma _{{\beta \sigma }}^{{\mu }}\Gamma _{{\alpha \rho }}^{{\rho }}-\Gamma _{{\sigma \rho }}^{{\mu }}\Gamma _{{\alpha \beta }}^{{\rho }}-\Gamma _{{\alpha \beta }}^{{\mu }}\Gamma _{{\sigma \rho }}^{{\rho }}) \\
{\displaystyle +g^{\alpha \beta }g^{\sigma \rho }(\Gamma _{\alpha \sigma }^{\mu }\Gamma _{\beta \rho }^{\nu }-\Gamma _{\alpha \beta }^{\mu }\Gamma _{\sigma \rho }^{\nu }))}+g^{{\alpha \beta }}g^{{\sigma \rho }}(\Gamma _{{\alpha \sigma }}^{{\mu }}\Gamma _{{\beta \rho }}^{{\nu }}-\Gamma _{{\alpha \beta }}^{{\mu }}\Gamma _{{\sigma \rho }}^{{\nu }})) ~.
\end{multline}
(See Appendix F of \cite{ShibataBook} or Chapter 20 of MTW \cite{misner1973gravitation} for a more thorough description.) This allows one to define a conservation law from the relation
\begin{equation}
	\partial_\mu (-g (T^{\mu\nu} + t_{LL}^{\mu\nu}) ) = 0 ~.
\end{equation}
It has the advantage over the Einstein pseudotensor that it is index symmetric, making it more appropriate for defining a global angular momentum of the spacetime.
Note that the conservation law contains the determinant of the metric $g$ and not the square root, which means that the resulting volume integrals are integrals of scalar densities of weight 1 rather than scalars. However, in cases where the asymptotic behaviour of the spacetime is flat with determinant 1, the volume integral of the charges again correspond at infinity to the ADM charges.\footnote{In the case where the determinant reaches a constant instead (e.g. in a cosmology with a non-unity scale factor) the charges differ by $\sqrt{g}_{r \to \infty}$ \cite{Landau:101808}.}
A second point to note is that the energy current must now be defined by $J^\mu = (-g) T^{\mu\nu} \delta^t_\nu$, that is $\tilde \zeta_\mu = (1,0,0,0)$ rather than $\tilde \zeta^\mu = (1,0,0,0)$. This implies that the conservation of charges in the normal way is associated with the case where the normal direction $n^\mu$ is a Killing symmetry rather than the time-like direction as in the Einstein pseudotensor description. As discussed above, in most numerical simulations, the stationary state obtained corresponds to the metric being constant in time, not in the normal direction $n^\mu$. Nevertheless one can construct the equivalent expressions in this formalism and use them in the same way, for example in the case of energy conservation one obtains that 
\begin{equation}
	\partial_t \left( \int_{\Sigma} d^3 x ~ \mathcal{Q} \right) = - \int_{\partial\Sigma} d^2 x ~ \mathcal{F} + \int_\Sigma d^3 x ~ \mathscr{S} ~,
	\label{eq:nconserveLL}
\end{equation}
with the charge
\begin{equation}
	\mathcal{Q} = (-g) ~ T^{00} =  \gamma \rho
\end{equation}
and flux (using the flat space normal direction $s_i$)
\begin{equation}
    \mathcal{F} = (-g) ~ T^{i0} s_i = \gamma (\alpha S^i - \beta^i \rho) s_i ~.
\end{equation}
The source is now
\begin{equation}
    \mathscr{S} = (-g) \left( T^{\mu\nu} ~^{(4)}\Gamma^t_{\mu \nu} - T^{\mu t} ~^{(4)}\Gamma^\nu_{\mu \nu} \right)
    = \gamma \left[ \rho \left( \partial_i \beta^i - \alpha K - \frac{\beta^i}{\alpha} \beta^j K_{ij}  \right) 
    + \alpha K^{ij} S_{ij} 
    + S^i (\alpha \Gamma^j_{ij} - \partial_i \alpha) ~.   \right] \label{eq:EsourceLL}
\end{equation}
Note the additional Christoffel symbol resulting from the scalar density weight.
The behaviour of the different formalisms is compared in Fig. \ref{fig-oscillaton} for the case of an oscillaton (as discussed in Sec. \ref{sec-applications}).

To summarise - assuming that we have an isolated system in an asymptotically flat region, we have proposed splitting the contributions to the ADM mass into a matter and a curvature component, and noted that by tracking the matter contribution we can infer the changes to the curvature contribution. However, as set out in the introduction, such a split is gauge dependent - the total ADM charges are the only unambiguous physical observables.
In a more general non asymptotically flat dynamical spacetime, the problem is even worse, since the total itself is dependent on both the gauge and the formalism chosen (ie, which curvature pseudotensor we define). So is it in fact useful to calculate such a split?

We would argue yes - it is still useful to track the volume integrals of the energy and source terms in order to gain an intuition about the transfer of energy between the spacetime and the matter, although one should take care not to make quantitative conclusions without considering the gauge dependence. In Sec. \ref{sec-applications}. we will give several simple examples that illustrate how this approach can be informative. If nothing else, the reconciliation of the matter charges is a useful check on the accuracy of the simulation, and often helps to identify issues such as boundary reflections, underresolved horizons and general code bugs.

\section{Applications}
\label{sec-applications}

In this section we discuss several specific simple examples that give an intuition for the continuity equation in general spacetimes, illustrate some of the points made above, and highlight its potential applications in numerical simulations of strong gravity environments. 

\subsection{Dynamical friction: the limit of small backreaction}

Dynamical friction is the gravitational drag force on an object due to the build up of an overdensity of matter behind it. A classic example is a black hole moving through a fluid and building up a gravitational ``wake'' of particles behind it. 
In the Newtonian description of such a system, the dynamical friction force $F^i$ is simply the integral of the gravitational force of each fluid element acting on the object, that is
\begin{equation}
    F_i = \int \frac{GM \rho}{r^2} s_i dV ~, \label{eq:newtonDF}
\end{equation}
with $s_i = \frac{x^i}{r}$ defining the projection of the force in the $i$-direction.
In the fully relativistic case the concept of a gravitational force is less straightforward to define - gravity is not a force but a curvature of spacetime - so one must instead define the force as the rate of change of the ADM momentum of the spacetime $F_i = \partial_t P_i$. This is assumed to be measured by some observer in an asymptotically flat region where their proper time $\tau = t$. (Strictly ADM charges are defined at infinity, where the fluxes should be zero, but the relations below hold approximately at a finite radius. To be clear we will write these approximate ADM charges as $\tilde P_\mu$.) Assuming the fluid has reached a steady state, we can get the force directly by using the surface integral equivalent to Eq. \eqref{eq:newtonDF}
\begin{equation}
    F_i = \int T^j_i dS_j = \partial_t \tilde P_i ~, \label{eq:newtonDF1}
\end{equation}
where the last equality assumes that the surface integral is evaluated in the asymptotically flat region, such that the result will hold also in GR.
One issue with this is that what we really want is the force on the BH due to the fluid, that is given that
\begin{equation}
	\tilde P_i = P_i^{curvature} + P_i^{matter} ~, \label{eq:split}
\end{equation}
we want the part associated with $\partial_t P_i^{curvature}$ only. If the matter has reached a steady state $\partial_t P_i^{matter} = 0$ then Eq. \eqref{eq:newtonDF1} gives the correct result. However, in a dynamical simulation often the matter takes time to reach a steady state around the BH, and it is useful to track the quantity specifically associated with the curvature in the period leading up to the stationary state being reached.
One usually finds that where a steady state is eventually reached, this term settles to a constant value well before the surface integral of the flux does so, which can reduce the time to solution for states that take a long time to settle. 

Consider a case where the stress energy of the fluid is small $T_{\mu\nu} \sim O(\epsilon)$, then to zeroth order in $\epsilon$ the spacetime is described by a vacuum solution $g_{\mu\nu}^{(0)}$. At first order in $\epsilon$ the propagation of the stress energy tensor is described by its evolution on the vacuum background
\begin{equation}
	\nabla_\rho(g_{\mu\nu}^{(0)}) T^{\rho \sigma} = 0 ~.
\end{equation}
There will be a resulting first order correction to the metric $\delta g_{\mu\nu} \sim O(\epsilon)$, which will have the effect of changing the energy-momentum of the spacetime curvature $P_\mu^{curvature}$ - it is this that results in the gravitational ``force'' on the BH. This correction will affect the propagation of the stress energy tensor, but only at higher orders in $\epsilon$, so we can neglect this backreaction effect, along with the gravitational wave flux $t^{GW}_{\mu\nu} \sim O(\epsilon^2)$. 
Taking care to include the effect of the flux of the matter across the BH horizon $\partial\Sigma_{BH}$ (which we take to contribute to $P_i^{curvature}$ for the BH), we find that to first order in $\epsilon$
\begin{equation}
    F_i \equiv \partial_t P_i^{curvature} = - \int_{\Sigma - \Sigma_{BH}} dV ~ \mathcal{S}_i - \int_{\partial\Sigma_{BH}} dS ~ \alpha N_j T^j_i ~.
    \label{eq:dPBHdt}
\end{equation}

\begin{figure}
\centering
\includegraphics[width=0.40\textwidth]{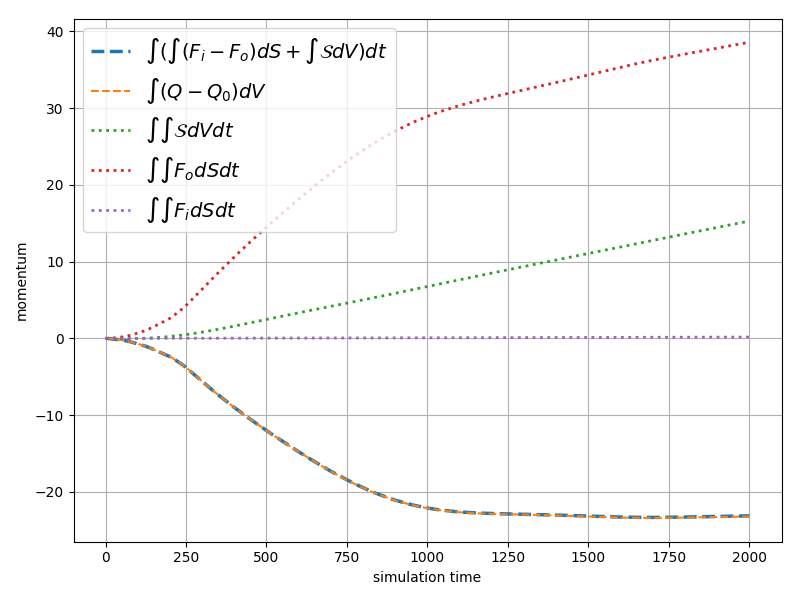}
\caption{\textit{Illustration of the agreement between the change in the total matter Momentum, and the time integrated Flux and the Source terms for the case of a BH moving through a scalar field. We see that the flux into the BH $F_i$ is negligible throughout. The physical impact of the source $\mathcal{S}$ is to change the curvature of the BH, reducing its momentum and leading to the effective drag force that we identify with the Newtonian idea of dynamical friction. Note that the source term settles to a constant value (as shown by the constant gradient in the figure) at around $t \sim 400$, whereas the outer flux $F_o$ does not settle to a constant value until $t \sim 1100$. Once a steady state is reached, the net momentum flux through the outer surface $F_o$ only contributes to slowing the BH via $\mathcal{S}$, and the cloud momentum $Q$ saturates to a roughly constant value.}}
\label{fig-df}
\end{figure}

A few comments on this result:
\begin{itemize}
	\item The force is composed of two contributions, the first being the dynamical friction effect of the cloud of matter behind the BH, and the second relating to accretion of the fluid's momentum onto the BH. 
	\item The split between these components is highly gauge dependent - in a Schwarzschild type metric clearly any flux measured at the horizon will be exactly zero, unlike in a horizon penetrating gauge.
	\item As discussed above, the source $\mathcal{S}_i$ quantifies the transfer of momentum between matter and curvature, which is the relativistic analogy of the Newtonian gravitational force being equal and opposite between two bodies.
	\item One can easily show that Eq. \eqref{eq:Psource} for $\mathcal{S}_i$ reduces to Eq. \eqref{eq:newtonDF1} in the Newtonian limit. 
\end{itemize} 

The agreement in the quantities extracted and their evolution over time is illustrated in Fig. \ref{fig-df}, coming from simulations carried out in \cite{Traykova:2021dua}.

\subsection{Oscillatons: quasi stable dynamical spacetime}

Quasi-stable localised configurations of massive, real scalar fields, called oscillatons \cite{Seidel:1991zh} balance the gravitational tendency of the field to collapse under gravity with its tendency to disperse due to gradient pressure. These are long lived localised solitons - a roughly gaussian-looking profile in the field that oscillates in time. Unlike the complex scalar case, the real scalar cannot support a time invariant stress-energy tensor and so both the metric and the field undergo oscillations in time. On each oscillation energy is transferred between the field and the spacetime curvature, and back again.

\begin{figure}[h]
\centering
\includegraphics[width=0.40\textwidth]{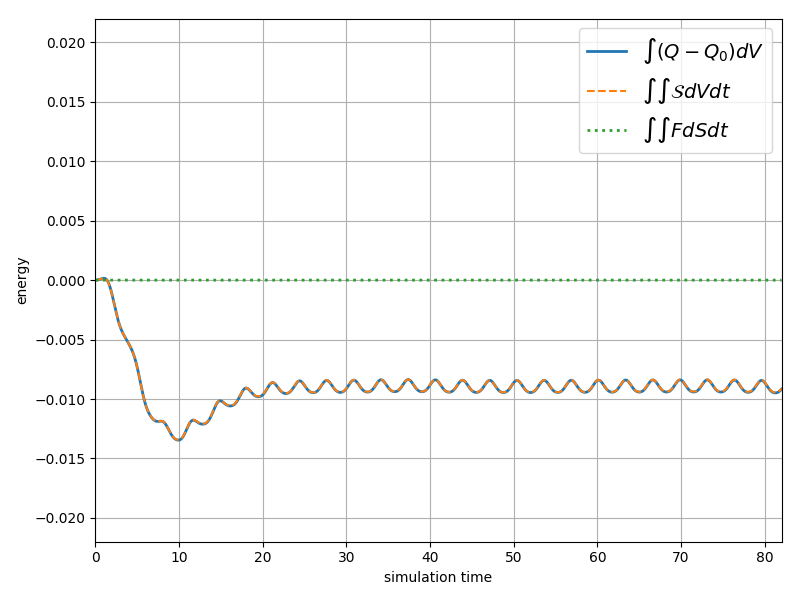}
\includegraphics[width=0.40\textwidth]{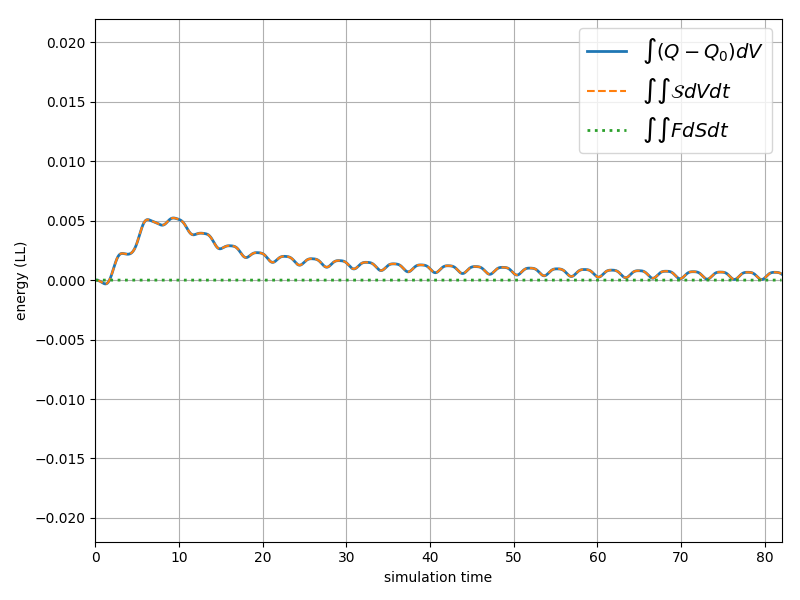}
\caption{\textit{Illustration of the agreement between the change in matter Energy, and the time integrated Flux and the Source terms for a simulation of an oscillaton, for the Einstein pseudotensor (left) and Landau Lifshitz pseudotensor (right) formulations. In both cases there is no flux $F$ out of the volume, and the change in the mass of the matter present is accounted for by the source term $\mathcal{S}$ quantifying the transfer between the curvature and the matter energy. As discussed in the text, the initial transient is related to the initial data adjusting dynamically into the simulation gauge, whereas the later steady state oscillations are physical transfers of the energy between the spacetime and the matter.}}
\label{fig-oscillaton}
\end{figure}

Due to the spherical symmetry of these solutions, no gravitational radiation is emitted, but scalar radiation is emitted on extremely long timescales, resulting in their eventual decay. On the timescales of an NR simulation, such a flux is negligible and so the initial ADM mass of the oscillaton is roughly conserved. The transfer of energy between the field and the metric can be tracked by considering Eq. \eqref{eq:Esource} for the matter energy source, and the result is plotted in Fig. \ref{fig-oscillaton} for the formalism based on both the Einstein and the Landau Lifshitz pseudotensors.

One can see for both cases that there is an initial transient in the source, which is unphysical, and relates solely to the the solution settling into its steady state in the dynamical puncture gauge used in the simulations, which is different from the areal polar gauge in which the initial data are derived. Eventually the source and mass of the matter settle into a steady oscillating state. Naively we might say that the total energy in the matter has decreased - but this cannot have happened since, as seen in the plots, there has been no flux out of the surface of the enclosing volume. In fact, since we have a new gauge and thus a new slicing of the 4D spacetime, what has changed is simply the split of what we denote to be matter energy and curvature energy. This illustrates clearly the gauge dependence of the split and the danger of interpreting such results too literally. However, once a quasi stable state is reached in the chosen gauge, the split is reasonably well defined and does tell us something physical about the oscillations of the energy between curvature and matter.

\subsection{FLRW spacetime expansion: no asymptotically flat space}

Another interesting illustration of the ideas above comes from their application to a spatially flat FLRW spacetime, in which there is no asymptotically flat space and no time-like Killing vector. 

One can show that due to the spatial flatness and isotropy, the total energy charge $E = -g(T^0_0 + t^0_0)$ in any spatial volume is zero, and any volume traced out by the comoving observers does not have a flux though its boundary. 
However, whilst the total energy is fixed (at zero), the energy content in the matter and the spacetime curvature can change over time, leading to the well known dependence of the behaviour of cosmological spacetimes on matter content.

Working in the comoving gauge with time defined by the proper time of the comoving observers,
\begin{equation}
	ds^2 = -dt^2 + a^2(t) \delta_{ij} dx^i dx^j ~,
\end{equation}
our continuity equation for the matter energy in a unit coordinate spatial volume (with the source Eq. \eqref{eq:Esource}) tells us that
\begin{equation}
	\partial_t (a^3 \rho) = a^3 (\frac{1}{2} S^{ij} \partial_t \gamma_{ij}) = -3a^2 \dot{a} P ~. \label{eq-frwcontinuity} 
\end{equation}
Here $P$ and $\rho$ are the pressure and energy density of the perfect fluid matter. This is the usual continuity equation of cosmological expansion, but written in a way that highlights the fact that what breaks ``normal'' energy conservation is the presence of pressure in the fluid - dust-like fluids with $P=0$ maintain the same total mass-energy even when the volume expands (the particles just get more spread out)
\footnote{In this case the conserved current is the baryonic number $J^\mu = \rho_0 u^\mu$ where $\rho_0$ is the rest mass density - the only energy comes from the bare mass contributions.}.
A positive pressure will tend to transfer energy from the matter to the spacetime curvature (e.g. in the case of radiation), whereas a negative pressure (e.g. a cosmological constant) will do the reverse.

We note that the requirement that $E = 0$ on each spatial slice translates to the first Friedmann equation
\begin{equation}
	\int d^3 x \sqrt{-g} (T^0_0 + t^0_0) = 0 \quad \implies \quad H^2 = \frac{8\pi \rho}{3} ~, \label{eq-friedman}
\end{equation}
which highlights the additional information required for full knowledge of the total energy-momentum of the spacetime.
(The requirement that $\partial_t E = 0$ gives the second Friedmann equation, but the same information can be obtained from a combination of Eq. \eqref{eq-friedman} and Eq. \eqref{eq-frwcontinuity}).

\section{Discussion}
\label{Discussion}

In this note we have provided an overview of the source terms that break the conservation of energy and momentum for general matter evolution in curved spacetimes. We have provided the relevant expressions for these sources in the ADM picture, and discussed applications and potential issues for several examples in the context of strong gravity simulations. We hope that this provides a useful reference on the topic, in particular for new students of NR. 

\section*{Acknowledgments}

\noindent KC thanks Jean Alexandre and her GRChombo collaborators (\url{https://www.grchombo.org/#people}), in particular Jamie Bamber, Robin Croft, Bo-Xuan Ge, Thomas Helfer, Eugene Lim, Miren Radia and Dina Traykova, for useful discussions, and acknowledges funding from the European Research Council (ERC) under the European Union's Horizon 2020 research and innovation programme (grant agreement No 693024). 

The simulations presented in this paper used DiRAC resources under the projects ACSP218 and ACTP238. This work was performed using the Cambridge Service for Data Driven Discovery (CSD3), part of which is operated by the University of Cambridge Research Computing on behalf of the STFC DiRAC HPC Facility (www.dirac.ac.uk). The DiRAC component of CSD3 was funded by BEIS capital funding via STFC capital grants ST/P002307/1 and ST/R002452/1 and STFC operations grant ST/R00689X/1. In addition used the DiRAC at Durham facility managed by the Institute for Computational Cosmology on behalf of the STFC DiRAC HPC Facility (www.dirac.ac.uk). The equipment was funded by BEIS capital funding via STFC capital grants ST/P002293/1 and ST/R002371/1, Durham University and STFC operations grant ST/R000832/1. DiRAC is part of the National e-Infrastructure.

\newpage 

\appendix

\section{Appendix: 4 dimensional Gauss Law in a general ADM spacetime}
\label{app:4DGauss}

The simplest way of applying Gauss's Law to a 4D spacetime is to start with a conservation law manipulated into a form in which it is expressed with partial derivatives only
\begin{equation}
	\partial_\mu \tilde{J}^\mu = 0 ~.
\end{equation}
For example, this can be achieved by defining $\tilde{J}^\mu = \sqrt{-g} J^\mu = \sqrt{-g} ~ T^\mu_\nu \zeta^\nu$ with $\zeta^\nu$ a Killing vector of the spacetime. Then one can write
\begin{equation}
	\partial_t \tilde{J}^0 = - \partial_i \tilde{J}^i
\end{equation}
and integrate both sides over a 3D spatial coordinate volume $\Sigma$ bounded by the 2D surface $\partial\Sigma$ - in which case one can directly apply Gauss theorem in 3D to find that
\begin{equation}
	\partial_t \int_\Sigma \tilde{J}^0 dV = - \int_{\partial \Sigma} \tilde{J}^i dS^f_i ~,
\end{equation}
where $dS^f_i$ is the coordinate surface integral element (ie, it does not contain factors of the metric, unlike the proper physical surface element $dS_i$), and we assume that the spatial volume $\Sigma$ is a constant coordinate volume with respect to time. For example in the case of the energy current for the coordinate observers, where $\zeta^\nu = \delta^\nu_0$, we obtain
\begin{equation}
	\partial_t \int_V \alpha T^0_0 ~  \sqrt{\gamma} dV = - \int_{\partial V} \alpha T^i_0 ~ \sqrt{\sigma}  dS_i \label{eq-basic-conserve} ~,
\end{equation}
where we have expanded out the$\sqrt{-g}$ factors in terms of the lapse and spatial metric to make the volume elements for the integrals physically meaningful. From this we can identify directly the conserved charge as $\alpha T^0_0$ and the flux as $\alpha T^i_0$.

A couple of implementation notes:
\begin{itemize}
\item Don't forget that $J^\mu$ is a four vector with non zero time-like components, and not a spatial 3-vector, so one cannot raise and lower the indices on $J^i$ with the spatial metric. In particular, this means that $J^i dS_i =  T^i_0 dS_i \neq  T_{i0} dS^i = J_i dS^i$.
\item In practise it is often useful to use the form with the flat surface integral element as this avoids the need to appropriately normalise $dS_i$ and calculate explicitly the induced metric on the surface $\sigma$. One can simply use the fact that $\sqrt{\sigma} N_i = \sqrt{\gamma} s_i$ where $s_i$ is a unit vector in the coordinate basis, i.e. $\delta^{ij} s_i s_j=1$.
\end{itemize}

\begin{figure}[h]
\centering
\includegraphics[width=0.5\textwidth]{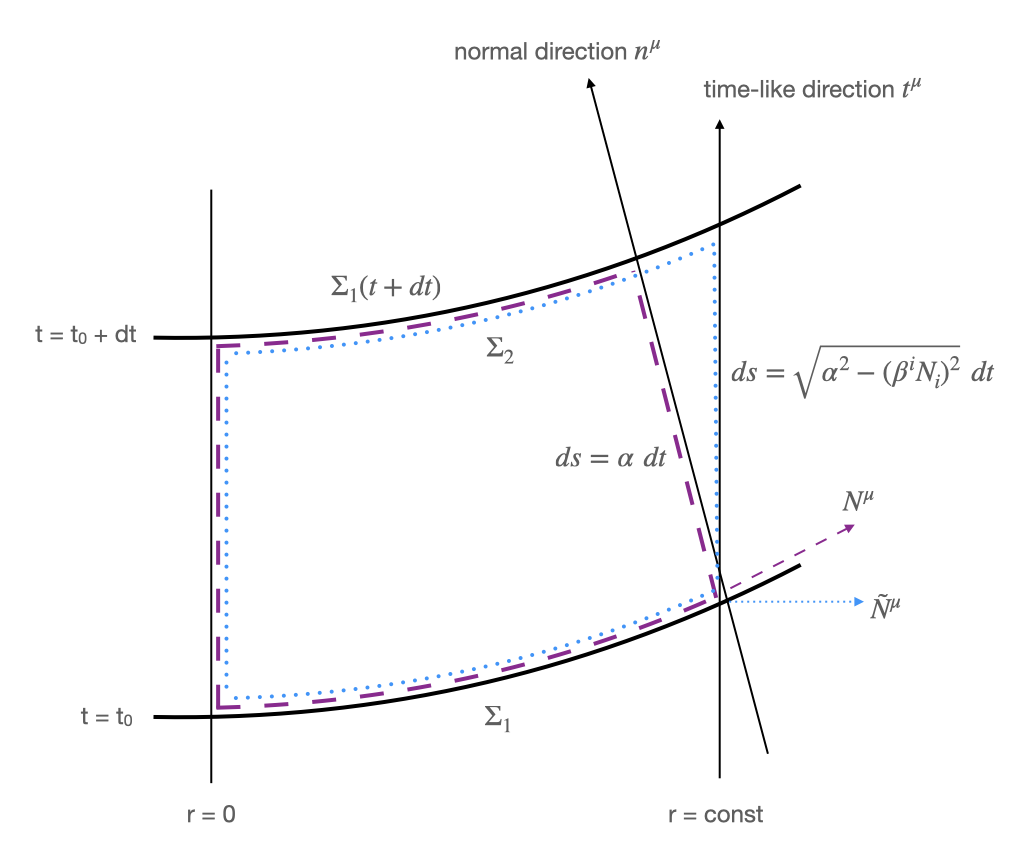}
\caption{\textit{Illustration of the ADM slicing of the 4D manifold $\mathcal{M}$ into 3D hypersurfaces $\Sigma$ bounded by the closed 2D surfaces $\partial\Sigma$ and separated by a small coordinate time interval $dt$. Volume 1 (as discussed in the text) is the dashed volume in purple and Volume 2 is the dotted volume in blue. The other 2 spatial dimensions are suppressed and we assume for illustration purposes that the spatial volumes are defined by spheres of constant coordinate radius $r$.}}
\label{fig-slices}
\end{figure}

Despite the elegance of this method, it is quite an instructive exercise to apply Gauss theorem directly to the 4D spacetime volume, particularly in the context of the ADM decomposition where we may have a non trivial shift and lapse, and convince ourselves that all the relevant metric factors work out and the four vectors reduce appropriately 3D spatial vectors. Below we derive explicitly the continuity equation using Gauss Law on the 4D volume in the ADM decomposition. The set up is illustrated in Fig. \ref{fig-slices}. This is adapted from section 3.5 of \cite{baumgarte_shapiro_2010}, but we will make some of the choices regarding volumes more explicit. 

Our starting point is the 4D volume integral written in covariant form
\begin{equation}
	\int d^4 x ~ \sqrt{-g} ~ \nabla_\mu J^\mu = 0 ~.
\end{equation}
We assume that we have a conserved current $J^\mu$ such that $\nabla_\mu J^\mu=0$. Note that $J^\mu = T^\mu_\nu \zeta^\nu$ is now properly a vector, not a tensor density like $\tilde{J}^\mu$ was.

In the ADM decomposition there are two natural 4D volumes we could consider, Volume 1 is the volume between the slice $\Sigma_1$ and $\Sigma_1(t+dt)$, defined by the congruence of the normal observers from one time slice to the next. Volume 2 is between $\Sigma_1$ and $\Sigma_2$, which have a constant coordinate volume in the spatial hyperslices, and are thus defined by the congruence of the time like observers. We will consider both in turn and show that the results agree - the considerations in each case are useful reminders of the physical properties of the slicing.

For Volume 1:
\begin{itemize}
\item The top and bottom spatial hyperslices, $\Sigma_1$ and $\Sigma_1(t+dt)$, are two 3D slices of constant $t$. They have normals $\pm n^\mu$ and the related metric determinant is $\gamma$. 
\item The spatial hyperslices are each bounded by a 2D surface $\partial\Sigma$ on which the induced metric has the determinant $\sigma$.
\item The sides of the volume trace the normal lines between the two slices. The normal to these sides lie wholly in the spatial surface, $n^\mu N_\mu = 0$, thus $N^\mu = (0,N^i)$, with the direction of $N^i$ defined via the 3D covector $s_i = \partial_i r$ (so $s_i= (1,0,0)$ in spherical coordinates or $x^i/r$ in cartesian ones) and normalised such that $g_{\mu\nu} N^\mu N^\nu = \gamma_{ij} N^i N^j = 1$. Thus explicitly,
\begin{equation}
	N^i = \frac{\gamma^{ij} s_j}{\sqrt{\gamma^{ij} s_i s_j}} ~.
\end{equation}
Note that we find from lowering the indices on $N^\mu$ that $N_0 = \beta^i N_i$.
\item The separation between the two spatial hypersurfaces in the normal $n^\mu$ direction has proper length  $ds = \alpha dt$.
\end{itemize}

For Volume 2:
\begin{itemize}
\item The top and bottom slices, $\Sigma_1$ and $\Sigma_2$, are two (3D) slices of constant $t$. They have normals $\pm n^\mu$ and the metric determinant is $\gamma$.
\item The spatial hyperslices are each bounded by a 2D surface $\partial\Sigma'$, which is a surface of constant $r$ and metric determinant $\sigma'$. (These coincide with $\partial\Sigma$ and $\sigma$ on $\Sigma_1$ but not in the rest of the volume.)
\item The sides of the volume trace the time vector $t^\mu = (1,0,0,0)$ - giving a surface of constant radial coordinate $r$.
\item The normal to the sides of the volume is the normal to the surface of constant $r$, thus it has a direction defined via the 4D covector $s_\mu = \partial_\mu r$ (so $s_\mu= (0,1,0,0)$ in spherical coords) and normalised such that $g_{\mu\nu} \tilde N^\mu \tilde N^\nu = 1$. One can show that this leads to the relation
\begin{equation}
	\tilde N_i = \frac{\alpha N_i}{\sqrt{\alpha^2 - (\beta^i N_i)^2}} \label{eqn-normals} ~.
\end{equation}
\item The separation between the two spatial hypersurfaces in the time-like ($t^\mu = (1,0,0,0)$) direction has proper length  $ds = (\alpha^2 - (\beta^i N_i)^2)^{1/2} dt$. Note that the minus sign here is important - it tells us that if the direction is null, the proper distance is zero. This ensures that when we integrate the flux across it, if the location of the boundary is moving outwards at the speed of light, the flux crossing it will be zero as expected (even if there is a non zero flux in the direction of the normal vector).
\end{itemize}

Finally, note that in a 4D Lorentzian spacetime a minus sign enters for space-like boundaries but not for time-like ones in Gauss's Law.

Starting with Volume 1, we convert the 4D volume integral to a sum of 3D surface integrals
\begin{equation}
\int d^3x (\mathcal{N}^\mu J_\mu)  = 
- \int_{\Sigma_1(t+dt)} d^3 x \sqrt{\gamma} ~ (n_\mu J^\mu) 
+ \int_{\Sigma_1} d^3 x \sqrt{\gamma} ~ (n_\mu J^\mu) 
+ \int_t^{t+dt} d\tilde t \int_{\partial\Sigma} d^2 x ~ \alpha \sqrt{\sigma} ~ (N_\mu J^\mu) 
= 0 ~.
\label{eqn-balance}
\end{equation}

We want to consider the limit of this volume when $dt$ is small, but we need to take care about the effect of the changing volume of the slices. In particular:
\begin{align}
	\partial_t \int_{\Sigma_1} d^3 x \sqrt{\gamma} ~  (n_\mu J^\mu) &= \lim_ {dt \rightarrow 0}  \frac{1}{dt} \left( \int_{\Sigma_2} d^3 x \sqrt{\gamma} ~ (n_\mu J^\mu)  - \int_{\Sigma_1} d^3 x \sqrt{\gamma} ~  (n_\mu J^\mu) \right) \\		
	&= \lim_ {dt \rightarrow 0}  \frac{1}{dt} \left( \int_{\Sigma_1(t+dt)} d^3 x \sqrt{\gamma} ~ (n_\mu J^\mu)  - \int_{\Sigma_1} d^3 x \sqrt{\gamma} ~  (n_\mu J^\mu) \right) + \int_{\partial\Sigma_1} d^2 x  \sqrt{\sigma} (\beta^i N_i) (n_\mu J^\mu) ~.
\end{align}

This means that, dividing Eqn. \eqref{eqn-balance} by $dt$ and taking the limit as $dt \rightarrow 0$ we obtain
\begin{equation}
    0 = \partial_t \int_{\Sigma_1} d^3 x \sqrt{\gamma} ~  (n_\mu J^\mu) -  \int_{\partial\Sigma_1} d^2 x ~ \sqrt{\sigma} ~ ( \alpha N_\mu J^\mu + \beta^i N_i n_\mu J^\mu) ~.
\end{equation}
Thus, remembering that $N_0 = \beta^i N_i$, we have
\begin{equation}
    \partial_t \int_{\Sigma_1} d^3 x \sqrt{\gamma} ~  (n_\mu J^\mu) =  \int_{\partial\Sigma_1} d^2 x ~ \sqrt{\sigma} ~ (\alpha ~ N_i J^i)  \label{eq:reconcile} ~,
\end{equation}
from which we can recover Eq. \eqref{eq-basic-conserve}.

Now for fun, let's check that we get the same result using Volume 2 instead.\footnote{Fun is, of course, an observer dependent quantity.} Now we have:
\begin{equation}
\int_{\Sigma_2} d^3 x \sqrt{\gamma} ~ (n_\mu J^\mu) 
- \int_{\Sigma_1} d^3 x \sqrt{\gamma} ~ (n_\mu J^\mu) 
- \int_t^{t+dt} d\tilde t \int_{\partial\Sigma'} d^2 x ~ ( (\alpha^2 - (\beta^i N_i)^2)^{1/2} ) \sqrt{\sigma'} ~ (\tilde N_\mu J^\mu) 
= 0 ~.
\label{eqn-balance2}
\end{equation}
Since now the coordinate volume of the spatial slices does not change, the first part is less fiddly. So we get directly that on the initial slice (where $\sigma = \sigma'$) 
\begin{equation}
    0 = \partial_t \int_{\Sigma_1} d^3 x \sqrt{\gamma} ~  (n_\mu J^\mu) -  \int_{\partial\Sigma_1} d^2 x ~ \sqrt{\sigma} ~ ( (\alpha^2 - (\beta^i N_i)^2)^{1/2} ) \tilde N_i J^i) ~.
\end{equation}
Using the relation between the two normals in Eqn. (\ref{eqn-normals}) we see that we recover the same result as in Eq. \eqref{eq:reconcile}.

\bibliography{NonConservation}

\end{document}